\documentclass[a4paper,11pt]{article}
\usepackage{pos}
\usepackage{braket}
\usepackage{array}  
\usepackage{xcolor}
\usepackage{orcidlink}

\newcolumntype{C}{>{$}c<{$}}

\title{Amplitude analysis of $\omega\pi^0$ photoproduction at GlueX}

\author*[a]{Kevin Scheuer \orcidlink{0009-0000-4604-9617}}

\onbehalf{for the GlueX collaboration}

\affiliation[a]{William \& Mary,\\
  300 Ukrop Way, Williamsburg, United States of America}

\emailAdd{kscheuer@wm.edu}

\FullConference{%
  The 10th International Conference on Quarks and Nuclear Physics,\\
  8-12 July 2024\\
  Facultat de Biologia Universitat de Barcelona, Barcelona, Spain
}

    \abstract{Excited meson states can often lie hidden within mass spectra beneath more dominant resonances, making it difficult to extract their physical properties. We can unveil these states through amplitude analysis, which disentangles the overlapping states via fits to their unique production and decay angular distributions. Understanding the light-meson spectrum is essential for confirming Lattice QCD predictions, especially in regards to the search for possible hybrid mesons. The Gluonic Excitation (GlueX) experiment at Jefferson Lab aids this search by studying the production of excited light mesons in $\gamma p$ interactions in the 8.2 - 8.8 GeV photon-beam energy range. We show in these proceedings preliminary results of an amplitude analysis of a large $\omega\pi^0$ dataset, concentrating on measuring the interference between the $b_1(1235)$ and an excited $J^{PC}=1^{--}$ vector resonance.}

\begin{document}
\maketitle

\section{Introduction} \label{sec:intro}
Hybrid mesons, or quark-antiquark states with an excited gluonic field configuration, are a particularly interesting area of research as their existence is not predicted by the quark model. Such states may populate areas of the light-quark meson spectrum like the high-mass $J^{PC}=1^{--}$ vector-meson region \cite{Dudek:2013yja}. An example of a well-known conventional quark-antiquark resonance is the $b_1(1235)$ with $J^{PC}=1^{+-}$, which decays dominantly to $\omega\pi$ and is well documented with several measurements of its mass, width, and its $\omega\pi$ $D$ to $S$-wave ratio \cite{ParticleDataGroup:2024cfk,E852:2002wsj}. Previous photoproduction measurements of $\gamma p\rightarrow \omega\pi^0 p$ indicate that the $b_1 \rightarrow \omega\pi^0$ amplitude should be dominant, however other amplitudes like excited $1^{--}$ vectors could contribute. An example is the $\omega\pi$ mode of the $\rho(1450)$, which does not have any measurements officially used for averages, fits, or limits in the PDG at this time \cite{ParticleDataGroup:2024cfk}. The most recent PDG-listed measurement is an $e^+e^-\rightarrow\omega\pi^0$ cross-section measurement based on a vector meson dominance model \cite{Achasov:2016zvn}. Photoproduction measurements of the $\omega\pi$ mode date back to 1984 and have limited precision \cite{OmegaPhoton:1983vde}. A more recent fit finds good agreement to the BESIII $e^+e^-\rightarrow\omega\pi^0$ cross-section data when using a set of various $1^{--}$ Breit-Wigner amplitudes, with strong interference effects \cite{BESIII:2020xmw}. The $\rho(1450)$ additionally is under debate for overlap with a possible separate $\rho(1250)$ state \cite{Hammoud:2020aqi}.

GlueX, a photoproduction experiment using a linearly polarized beam, is well suited to address the need for further measurements of the $b_1$ and excited vector mesons. The detector measures charged tracks and neutral particle showers over almost the entire solid angle surrounding the proton target. A complete description of the experimental setup can be found elsewhere \cite{GlueX:2020idb}. The first data-taking campaign has been completed and is labeled here as the ``GlueX Phase-I'' dataset, corresponding to an integrated luminosity of $\sim$125 pb$^{-1}$, resulting in data samples roughly 3 orders of magnitude greater than those from previous experiments \cite{OmegaPhoton:1983vde}. In these proceedings, we study exclusive events for the reaction $\gamma p \rightarrow \omega\pi^0$ where $\omega\rightarrow\pi^+\pi^-\pi^0$ in the beam energy range $8.2 < E_\gamma < 8.8$ GeV where the linear photon beam polarization is $\sim$35\%. The analysis is separated into 4 bins of the squared four-momentum transfer $-t$ from 0.1 to 0.9 GeV${}^2$, with roughly equal number of events in each bin, to study the production mechanism.  To focus on the potential interference between the well-known $b_1(1235)$ and a possible excited vector resonance, we perform the amplitude analysis in the $\omega\pi^0$ invariant mass region from 1.0 to 1.4 GeV. This reasonably excludes contributions from the possible $\rho_3(1690)$ or $\rho(1700)$ resonances, simplifying the set of amplitudes required to describe the data. 

\section{Amplitude Analysis Model}
We fit the measured intensity distribution with an amplitude model that takes into account the beam-photon polarization to determine the interfering contributions between waves with the quantum numbers reflectivity $\varepsilon$, total angular momentum $J$, parity $P$, orbital angular momentum $\ell$, and spin-projection $m$. The reflectivity quantum number is well described, and because we are limiting ourselves to $-t\leq 0.9$ GeV$^2$, has a direct relation to the naturality $\eta=P(-1)^J$ of the $t$-channel exchange particle \cite{Mathieu:2019fts}. The model
\begin{equation} \label{eq:pwa-model}
\begin{aligned}
    I(\Phi, \Omega, \Omega_H) = &(1-P_\gamma)\bigg[
        \Big|
            \sum_{i,m} [c_i]^{-}_{m} \Im (Z^i_m)
        \Big|^2
        + \Big|
            \sum_{i,m} [c_i]^{+}_{m} \Re (Z^i_m)
        \Big|^2
    \bigg]
    \\
    + &(1+P_\gamma)\bigg[
        \Big|
            \sum_{i,m} [c_i]^{+}_{m} \Im (Z^i_m)
        \Big|^2
        + \Big|
            \sum_{i,m} [c_i]^{-}_{m} \Re (Z^i_m)
        \Big|^2
    \bigg]
\end{aligned}
\end{equation}
used for vector-pseudoscalar processes was adapted from the two-pseudoscalar model from Ref.~\cite{Mathieu:2019fts} by accounting for the extra angular information due to the decay of the vector daughter. In Eq.(~\ref{eq:pwa-model}), the $c$ are the free complex production parameters, $i$ indexes every possible $J^P\ell$ combination, $P_\gamma$ is the polarization fraction, and 
\begin{equation}
    Z^i_m (\Phi, \Omega, \Omega_H) = \mathrm{e}^{-i\Phi} \sum_{\lambda_\omega=-1}^{+1} D_{m,\lambda_\omega}^{J_i\ast}(\Omega) \Big( 
        \sum_{l_i} \braket{J_i\lambda_\omega|\ell_i0,1\lambda_\omega}C_{\ell_i}
    \Big)
    D_{0,\lambda_\omega}^{1\ast}(\Omega_H) G_{\text{Dalitz}}
\end{equation}
contains the angular information. Here, $\Phi$ is the angle between the production plane and the beam photon's polarization vector, $\Omega$ describes the angular information of the vector particle $(\omega)$ in the rest frame of the $\omega\pi^0$ pair, and $\Omega_H$ the subsequent decay of $\omega\rightarrow \pi^+\pi^-\pi^0$ in the $\omega$ rest frame. 

The $G_{\text{Dalitz}}$ factor models the Dalitz distribution of the pions in the $\omega$ decay, with its parameters fixed to theoretical predictions from Ref.~\cite{JPAC:2020umo}. The fits use AmpTools, a library primarily built for performing unbinned extended maximum-likelihood fits using MINUIT \cite{matthew_shepherd_2024_10961168}.

\subsection{Verifying the Amplitude Model with Monte Carlo}

\begin{figure}
    \centering
    \includegraphics[width=0.5\linewidth,angle=270]{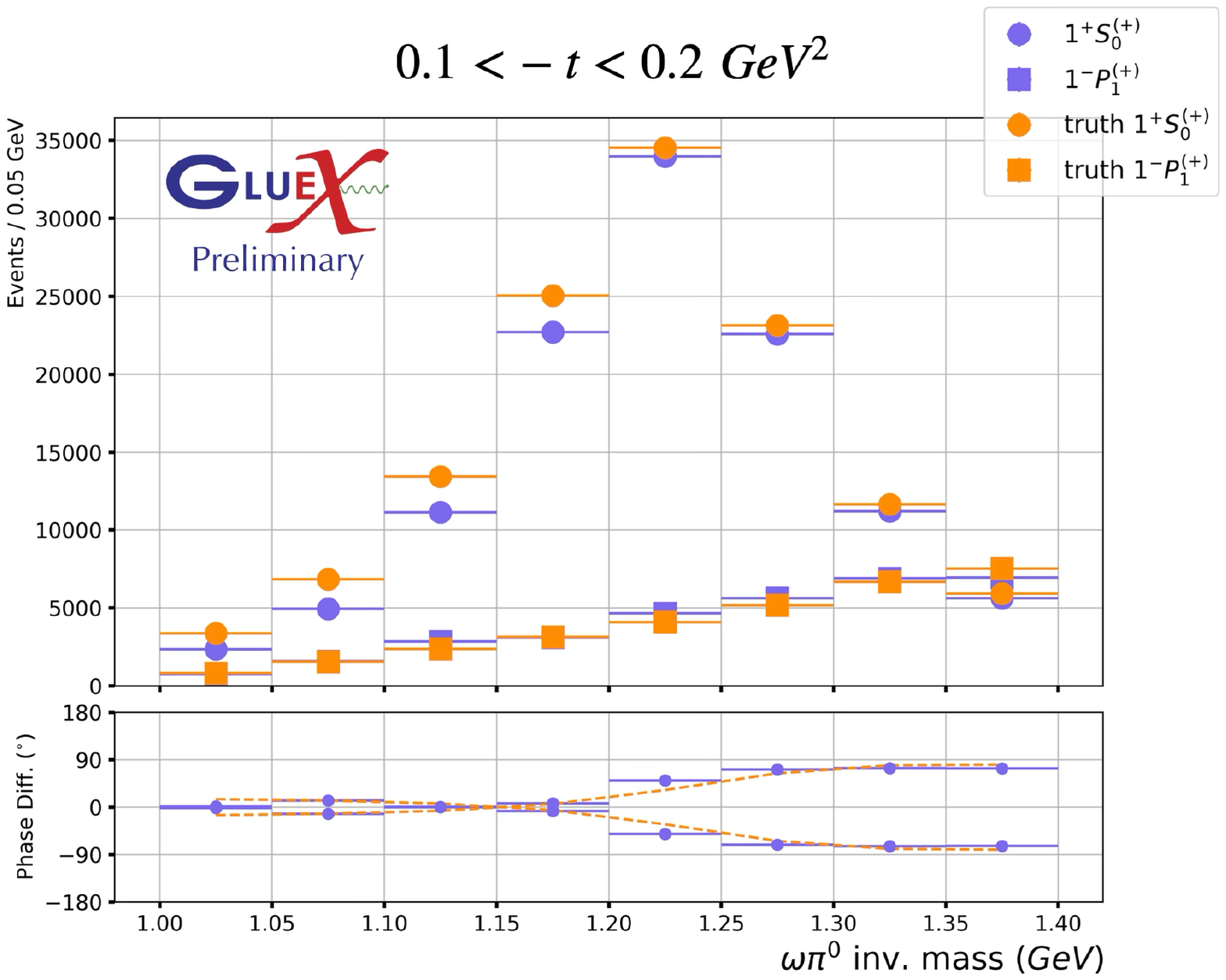}
    \caption{The two dominant $J^P \ell_m^{(\varepsilon)}$ waves that captured the $b_1(1235)$ and $\rho(1450)$ contributions generated by Monte Carlo simulation. Output from the \textcolor[HTML]{F09235}{mass-independent fit} are compared to their \textcolor[HTML]{7869E6}{true, originally generated input values}. The model inherently has an ambiguity in the sign of the phase difference, so both sets of solutions are plotted in each bin.}
    \label{fig:io_test}
\end{figure}

To ensure that the amplitude model can obtain unique results, we perform an ``Input-Output'' study in a limited $-t$ region. We obtain the input by fitting GlueX Phase-I $\omega\pi^0$ events with a mass-dependent version of the model in Eq.(~\ref{eq:pwa-model}), in which we effectively require that the data follow a Breit-Wigner lineshape. As discussed above, the limited mass region is well suited to study the interference between $b_1(1235)$ and $\rho(1450)$ resonances, and so they are included in the fit with their masses and widths fixed to their PDG values \cite{ParticleDataGroup:2024cfk}. 

Based on this fit result, we generate Monte Carlo simulated events and pass them through a GEANT4 simulation of the GlueX detector, and reconstruct them with the same procedure used for the data. The resulting detected events then serve as the "input" part of the study. We fit the mass-independent model (no Breit-Wigner lineshapes assumed) to the input data and compare the parameter values from this ``output'' fit result to the generated values. We restrict the waveset to the same waves as the generated data, i.e. $J^P\ell = \{1^+S,1^+D,1^-P\}$ with all 3 $m$-projections -1, 0, +1, and both reflectivities $\varepsilon=\pm1$ allowed. We observe that the $J^P \ell_m^{(\varepsilon)} = 1^+ S_0^{(+)}, 1^-P_1^{(+)}$ waves capture the majority of the $b_1$ and $\rho$ signals respectively, as shown in Figure~\ref{fig:io_test}. We also observe a significant phase motion across the analyzed mass range, which the mass-independent fit can successfully reproduce.

\section{Results}

\begin{figure}
    \centering
    \includegraphics[width=0.5\linewidth,angle=270]{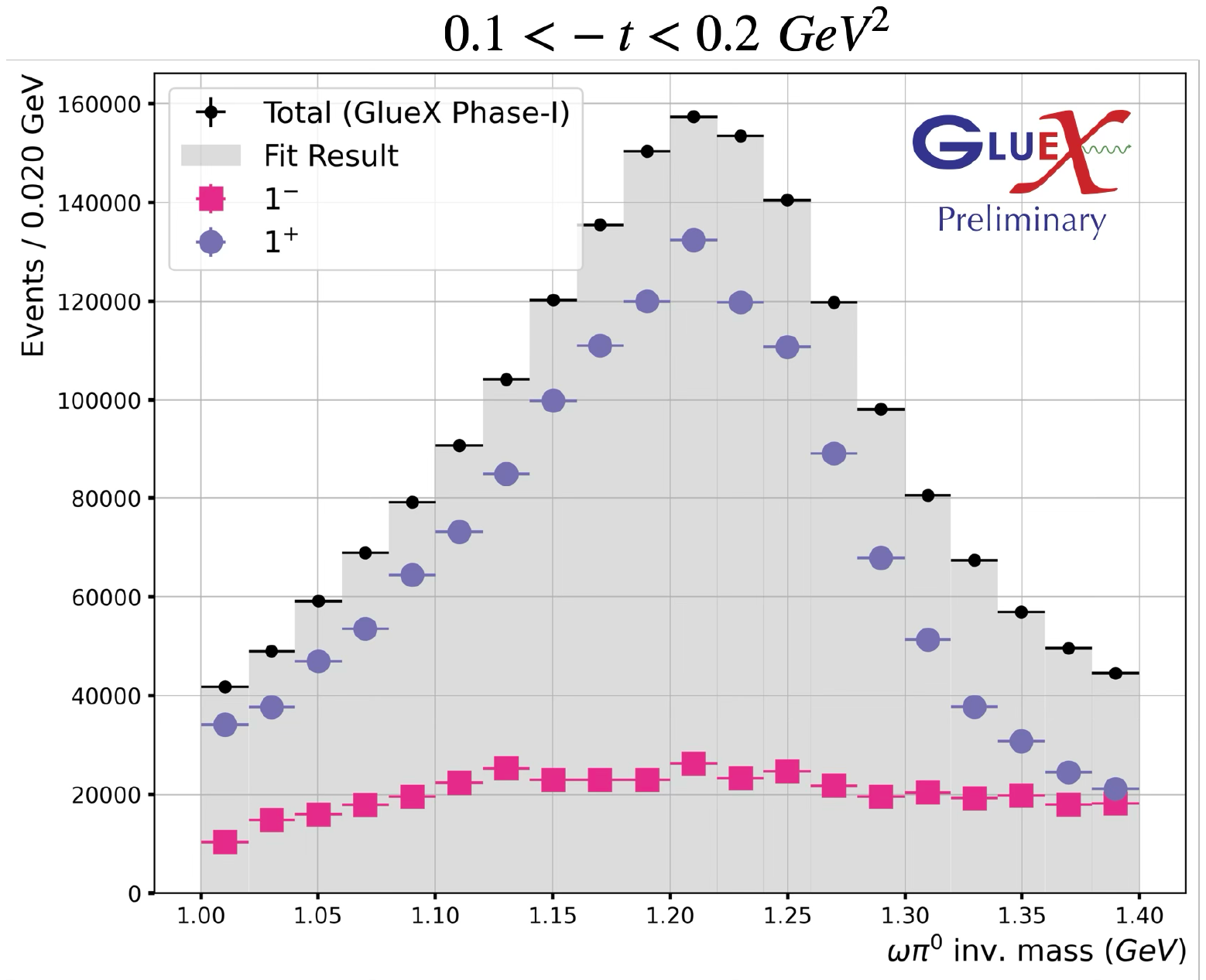}
    \caption{Preliminary results of a mass-independent fit to GlueX Phase-I data. The grey histogram, obtained by fitting the angular distribution of the events, successfully reproduces the intensity in each $\omega\pi^0$ invariant mass bin. We decompose the total intensity into its individual $J^P$ contributions, showing a clearly dominant $1^+$ signal consistent with a $b_1(1235)$. Note that errors, which are present but hidden by the markers, are purely statistical.}
    \label{fig:jp}
\end{figure}

Having verified the model's abilities to extract amplitudes reliably, we perform a mass-independent fit on the GlueX Phase-I dataset. The same waveset used in the input-output study is applied and Figure~\ref{fig:jp} shows the individual $J^P$ contributions obtained by coherently summing over the different $\ell$, $m$, and $\varepsilon$ waves belonging to each $J^P$ value. The fit clearly shows a dominant $1^+$ contribution consistent with the E852 results, and previous photoproduction results \cite{E852:2002wsj, OmegaPhoton:1983vde}. Note that uncertainties are purely statistical in all figures, and we plan to conduct systematic studies in future work. 

\begin{figure}
    \centering
    \includegraphics[width=0.5\linewidth,angle=270]{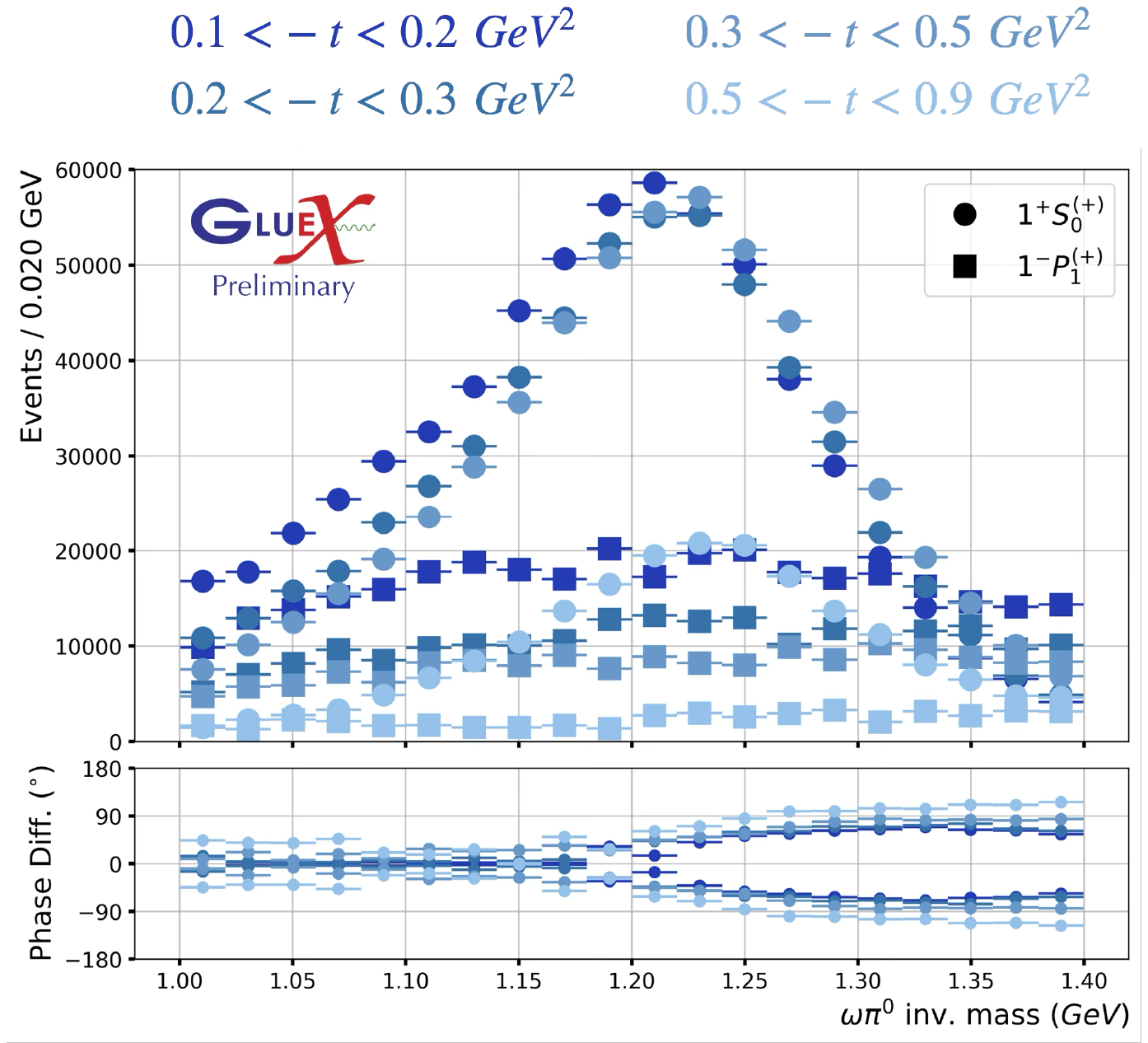}
    \caption{Preliminary results of a mass-independent amplitude analysis on GlueX Phase-I Data in the $\omega\pi^0$ channel. The shapes of the markers correspond to the wave and the colors to the $-t$ bin the fits were performed in. The phase difference of the two $J^P\ell_m^{(\varepsilon)}$ amplitudes predicted to primarily capture the $b_1(1235)$ and a $1^{--}$ vector resonance are shown. We observe phase motion across the mass bins that appears consistent between $-t$ bins. Note that uncertainties, which are present but covered by the markers, are purely statistical.}
    \label{fig:final}
\end{figure}

Figure~\ref{fig:final} shows preliminary results of the two dominant amplitudes in all four $-t$ bins. The phase motion seems to occur independent of $-t$. Additionally, the phase motion is smooth across the mass range even for the highest $-t$ bin whose $J^P\ell_m^{(\varepsilon)} = 1^-P_1^{(+)}$ wave intensity is quite small. These results are consistent with the findings of the input-output study and suggest that the $b_1(1235)$ is interfering with a $1^{--}$ vector amplitude. 

\section{Conclusions}
We present an amplitude analysis of the polarized photoproduction reaction $\gamma p\rightarrow \omega\pi^0 p$ to investigate the interference between the $b_1(1235)$ and possible $1^{--}$ vector resonances. Performing an input-output test using simulated Monte Carlo data, we verify that the $b_1$ and a vector meson, with their relative phase difference, are primarily captured by the $J^P\ell_m^{(\varepsilon)} = 1^+S_0^{(+)}$ and $1^{-}P_1^{(+)}$ amplitudes, respectively. We observe a phase motion independent of $-t$ in fits to data that hints at resonance interference activity occurring. This result utilizes one of the largest $\omega\pi^0$ photoproduction datasets to date and contributes towards the effort to understand the spectrum of light mesons. 
\newpage

\acknowledgments{
    Thanks to Justin Stevens, Edmundo Barriga, Amy Schertz, Matt Shepherd, Boris Grube, and all my other colleagues in the GlueX collaboration for their guidance and help in making this analysis possible. This work was supported by the U.S. Department of Energy, Office of Science, Office of Nuclear Physics under the award DE-SC0023978.
}

\bibliographystyle{JHEP}
\bibliography{refs}

\end{document}